\documentclass[showpacs,preprintnumbers,
amsmath,amssymb,aps,prd]{revtex4}
\usepackage{}
\usepackage{amsfonts}
\usepackage{amsfonts,graphics,epsfig,subfigure}

\begin{document}

\title{$P-V$ Criticality of Topological Black Holes in Lovelock-Born-Infeld Gravity}
\author{Jie-Xiong Mo $^{a,b}$ \footnote{mojiexiong@gmail.com}, Wen-Biao Liu $^a$ \footnote{wbliu@bnu.edu.cn (corresponding
author)}}

 \affiliation{$^a$ Department of Physics, Institute of Theoretical Physics, Beijing Normal University, Beijing, 100875, China\\
 $^b$ Institute of Theoretical Physics, Zhanjiang Normal University, Zhanjiang, 524048, China\\
  }

\begin{abstract}
  To understand the effect of third order Lovelock gravity, $P-V$ criticality of topological AdS black holes in Lovelock-Born-Infeld gravity is investigated. The thermodynamics is further explored with some more extensions and details than the former literature. A detailed analysis of the limit case $\beta\rightarrow\infty$ is performed for the seven-dimensional black holes. It is shown that for the spherical topology, $P-V$ criticality exists for both the uncharged and charged cases. Our results demonstrate again that the charge is not the indispensable condition of $P-V$ criticality. It may be attributed to the effect of higher derivative terms of curvature because similar phenomenon was also found for Gauss-Bonnet black holes. For $k=0$, there would be no $P-V$ criticality. Interesting findings occur in the case $k=-1$, in which positive solutions of critical points are found for both the uncharged and charged cases. However, the $P-v$ diagram is quite strange. To check whether these findings are physical, we give the analysis on the non-negative definiteness condition of entropy. It is shown that for any nontrivial value of $\alpha$, the entropy is always positive for any specific volume $v$. Since no $P-V$ criticality exists for $k=-1$ in Einstein gravity and Gauss-Bonnet gravity, we can relate our findings with the peculiar property of third order Lovelock gravity. The entropy in third order Lovelock gravity consists of extra terms which is absent in the Gauss-Bonnet black holes, which makes the critical points satisfy the constraint of non-negative definiteness condition of entropy. We also check the Gibbs free energy graph and the "swallow tail" behavior can be observed. Moreover, the effect of nonlinear electrodynamics is also included in our research.
\end{abstract}

\keywords{$P-V$ Criticality\;  Topological black holes\; Lovelock gravity}
 \pacs{04.70.Dy, 04.70.-s} \maketitle

\section{Introduction}
Gravity in higher dimensions has attained considerable attention with the development of string theory. Concerning the effect of string theory on gravitational physics, one may construct a low energy effective action which includes both the Einstein-Hilbert Lagrangian (as the first order term) and higher curvature terms. However, this approach may lead to field equations of fourth order and ghosts as well. This problem has been solved by a particular higher curvature gravity theory called Lovelock gravity~\cite{Lovelock}. The field equation in this gravity theory is only second order and the quantization of Lovelock gravity theory is free of ghosts~\cite{Boulware}. In this context, it is of interest to investigate both the black hole solutions and their thermodynamics in Lovelock gravity~\cite{Dehghani1}-\cite{Amirabi}. Moreover, it is natural to consider the nonlinear terms in the matter side of the action while accepting the nonlinear terms on the gravity side~\cite{Dehghani1}. Motivated by this, Ref.~\cite{Dehghani1} presented topological black hole solutions in Lovelock-Born-Infeld gravity. Both the thermodynamics of asymptotically flat black holes for $k=1$ and the thermodynamics of asymptotically AdS rotating black branes with flat horizon were detailedly investigated there. However, concerning the charged topological AdS black holes in Lovelock-Born-Infeld gravity, only the temperature was given in Ref.~\cite{Dehghani1}. Ref.~\cite{Decheng2} further studied their entropy and specific heat at constant charge. However, the expression of entropy seems incomplete for $k$ is missing. And the thermodynamics in the extended space needs to be further explored. Probing this issue is important because it is believed that the physics of black holes in higher dimensions is essential for one to understand a full theory of quantum gravity.

     As is well known, phase transition is a fascinating phenomenon in the classical thermodynamics. Over the past decades, phase transitions of black holes have aroused more and more attention. The pioneer phase transition research of AdS black holes can be traced back to the discovery of the famous Hawking-Page phase transition between the Schwarzschild AdS black hole and thermal AdS space \cite{Hawking2}. Recently, a revolution in this field is led by $P-V$ criticality research~\cite{Kubiznak}-\cite{Decheng} in the extended phase space. Kubiz\v{n}\'{a}k et al.~\cite{Kubiznak} perfectly completed the analogy between charged AdS black holes and the liquid-gas system first observed by Chamblin et al.~\cite{Chamblin1,Chamblin2}. The approach of treating the cosmological constant as thermodynamic pressure and its conjugate quantity as thermodynamic volume is essential with the increasing attention of considering the variation of the cosmological constant in the first law of black hole thermodynamics recently~\cite{Caldarelli}-\cite{Lu} .

Here, we would like to investigate the thermodynamics and phase transition of charged topological AdS black holes in Lovelock-Born-Infeld gravity in the extended phase space. Some related efforts have been made recently. $P-V$ criticality of both four-dimensional~\cite{Gunasekaran} and higher dimensional~\cite{Decheng} Born-Infeld AdS black holes have been investigated. A new parameter called Born-Infeld vacuum polarization was defined to be conjugated to the Born-Infeld parameter~\cite{Gunasekaran}. And it was argued that this quantity
is required for the consistency of both the first law of thermodynamics and the Smarr relation. Moreover, Cai et al.~\cite{Cai98} studied the $P-V$ criticality of Gauss-Bonnet AdS black holes. It was found that no $P-V$ criticality can be observed for Ricci flat and hyperbolic Gauss-Bonnet black holes. However, for the spherical case, $P-V$ criticality can be observed even when the charge is absent, implying that the charge may not be the indispensable factor for the existence of $P-V$ criticality. Such an interesting result motivates us to probe further the third order Lovelock gravity to explore whether it is a peculiar property due to the higher derivative terms of curvature. So we would mainly investigate their effects on the $P-V$ criticality. Moreover, we will probe the combined effects of higher derivative terms of curvature and the nonlinear electrodynamics.

 In Sec. \ref{Sec2}, the solutions of charged topological AdS black holes in Lovelock-Born-Infeld gravity will be briefly reviewed and their thermodynamics will be further investigated. In Sec. \ref{Sec3}, a detailed study will be carried out in the extended phase space for the limit case $\beta\rightarrow\infty$ so that we can concentrate on the effects of third order Lovelock gravity. In Sec. \ref{Sec4}, the effects of nonlinear electrodynamics will also be included. In the end, a brief conclusion will be drawn in Sec. \ref {Sec5}.
\section{Thermodynamics of charged topological black holes in Lovelock-Born-Infeld gravity}

\label{Sec2}
The action of third order Lovelock gravity with nonlinear Born-Infeld electromagnetic field is~\cite{Dehghani1}
\begin{equation}
I_{G}=\frac{1}{16\pi}\int d^{n+1}x\sqrt{-g}\big(-2\Lambda+\mathcal{L}_1+\alpha_2\mathcal{L}_2+\alpha_3\mathcal{L}_3+L(F)\big),\label{1}
\end{equation}%
where
\begin{eqnarray}
\mathcal{L}_1&=&R,\label{2}
\\
\mathcal{L}_2&=&R_{\mu\nu\gamma\delta}R^{\mu\nu\gamma\delta}-4R_{\mu\nu}R^{\mu\nu}+R^2, \label{3}
\\
\mathcal{L}_3&=&2R^{\mu\nu\sigma\kappa}R_{\sigma\kappa\rho\tau}R^{\rho\tau}_{\;\;\;\;\mu\nu}+8R^{\mu\nu}_{\;\;\;\;\sigma\rho}R^{\sigma\kappa}_{\;\;\;\;\nu\tau}R^{\rho\tau}_{\;\;\;\;\mu\kappa}+24R^{\mu\nu\sigma\kappa}R_{\sigma\kappa\nu\rho}R^{\rho}_{\;\;\mu} \nonumber
\\
&\,&+3RR^{\mu\nu\sigma\kappa}R_{\sigma\kappa\mu\nu}+24R^{\mu\nu\sigma\kappa}R_{\sigma\mu}R_{\kappa\nu}+16R^{\mu\nu}R_{\nu\sigma}R^{\sigma}_{\;\;\mu}-12RR^{\mu\nu}R_{\mu\nu}+R^3,\label{4}
\\
L(F)&=&4\beta^2\left(1-\sqrt{1+\frac{F^2}{2\beta^2}}\right). \label{5}
\end{eqnarray}%
In the above action, $\beta$, $\alpha_2$ and $\alpha_3$ are Born-Infeld parameter, the second and third order Lovelock coefficients respectively while $\mathcal{L}_1$, $\mathcal{L}_2$, $\mathcal{L}_3$ and $L(F)$ are Einstein-Hilbert, Gauss-Bonnet, the third order Lovelock and Born-Infeld Lagrangians respectively. Considering the case
\begin{eqnarray}
\alpha_2&=&\frac{\alpha}{(n-2)(n-3)}, \label{6}
\\
\alpha_3&=&\frac{\alpha^2}{72{n-2\choose 4}},\label{7}
\end{eqnarray}%
Ref.~\cite{Dehghani1} derived the $(n+1)$-dimensional static solution as
\begin{equation}
ds^2=-f(r)dt^2+\frac{dr^2}{f(r)}+r^2d\Omega^2, \label{8}
\end{equation}%
where
\begin{eqnarray}
f(r)&=&k+\frac{r^2}{\alpha}(1-g(r)^{1/3}),\label{9}\\
g(r)&=&1+\frac{3\alpha m}{r^n}-\frac{12\alpha \beta^2}{n(n-1)}\Big[1-\sqrt{1+\eta}-\frac{\Lambda}{2\beta^2}+\frac{(n-1)\eta}{(n-2)}\digamma(\eta)\Big],\label{10}\\
d\Omega^2&=&\left\{
\begin{array}[c]{l}
d\theta_1^2+\overset{n-1}{\underset{i=2}{{\textstyle\sum}}}\overset{i-1}{\underset{j=1}{{\textstyle\prod}}}\sin^2\theta_jd\theta^2_i\text{\ \ \ \ \ \ \ \ \ \ \ \ \ \ \ \ \ \ \ \ \ \ \ \ \ \ \ \ \ \ \ \ \ \ \ \ \ \ \ \ \ \ \ \ $k$=1}\\
d\theta_1^2+\sinh^2\theta_1d\theta_2^2+\sinh^2\theta_1\overset{n-1}{\underset{i=3}{{\textstyle\sum}}}\overset{i-1}{\underset{j=2}{{\textstyle\prod}}}\sin^2\theta_jd\theta^2_i\text{\ \ \ \ \ \ \ \ \ \ \ \ \ \ \ \ $k$=-1}\\
\overset{n-1}{\underset{i=1}{{\textstyle\sum}}}d\phi^2_i\text{\ \ \ \ \ \ \ \ \ \ \ \ \ \ \ \ \ \ \ \ \ \ \ \ \ \ \ \ \ \ \ \ \ \ \ \ \ \ \ \ \ \ \ \ \ \ \ \ \ \ \ \ \ \ \ \ \ \ \ \ \ \ \ \ \ $k$=0}
\end{array}
\right..
\label{11}
\end{eqnarray}%
$d\Omega^2$ denotes the line element of $(n-1)$-dimensional hypersurface with constant curvature $(n-1)(n-2)k$ and $\digamma(\eta)$ denotes the hypergeometric function as follow
\begin{equation}
\digamma(\eta)=\,_2F_1\Big(\Big[\frac{1}{2},\frac{n-2}{2n-2}\Big],\Big[\frac{3n-4}{2n-2}\Big],-\eta\Big), \label{12}
\end{equation}%
where
\begin{equation}
\eta=\frac{(n-1)(n-2)q^2}{2\beta^2r^{2n-2}}. \label{13}
\end{equation}%
The Hawking temperature has been derived in Ref.~\cite{Dehghani1} as
\begin{equation}
T=\frac{(n-1)k[3(n-2)r_+^4+3(n-4)k\alpha r_+^2+(n-6)k^2\alpha^2]+12r_+^6\beta^2(1-\sqrt{1+\eta_+}\,)-6\Lambda r_+^6}{12\pi(n-1)r_+(r_+^2+k\alpha)^2}.\label{14}
\end{equation}%
However, only the Hawking temperature is not enough to discuss the $P-V$ criticality in the extended phase space. So we would like to calculate other relevant quantities.

Solving the equation $f(r)=0$, one can obtain the parameter $m$ in terms of the horizon radius $r_+$ as
\begin{equation}
m=\frac{r_+^n}{3\alpha}\left\{-1+\frac{(r_+^2+k\alpha)^3}{r_+^6}+\frac{12\alpha\beta^2\left[1-\frac{\Lambda}{2\beta^2}-\sqrt{1+\eta}+\frac{(n-1)\digamma(\eta)\eta}{n-2}\right]}{n(n-1)}\right\}.\label{15}
\end{equation}%
Then the mass of $(n+1)$-dimensional topological AdS black holes can be derived as
\begin{equation}
M=\frac{(n-1)\Sigma_k}{16\pi}m=\frac{(n-1)\Sigma_k r_+^n}{48\pi\alpha}\left\{-1+\frac{(r_+^2+k\alpha)^3}{r_+^6}+\frac{12\alpha\beta^2\left[1-\frac{\Lambda}{2\beta^2}-\sqrt{1+\eta}+\frac{(n-1)\digamma(\eta)\eta}{n-2}\right]}{n(n-1)}\right\},\label{16}
\end{equation}%
where $\Sigma_k$ denotes the volume of the $(n-1)$-dimensional hypersurface mentioned above.

The entropy can be calculated as
\begin{equation}
S=\int^{r_+}_{0}\frac{1}{T}\left(\frac{\partial M}{\partial r_+}\right)dr=\frac{\Sigma_k(n-1)r_+^{n-5}}{4}\left(\frac{r_+^4}{n-1}+\frac{2kr_+^2\alpha}{n-3}+\frac{k^2\alpha^2}{n-5}\right).\label{17}
\end{equation}%
Note that the above integration is accomplished under the condition of $n>5$. For $n\leqslant5$ the integration is divergent. So in this paper, we would mainly investigate the case of
$n=6$, which corresponds to the seven-dimensional black holes. The third term of the entropy in Eq. (\ref{17}) does not appear in the expression of the entropy of Gauss-Bonnet black holes~\cite{Cai98}. Our result also extends the expression in Ref.~\cite{Decheng2} where $k$ was missing.

In the extended phase space, one may identify the pressure of the black hole as~\cite{Kubiznak}
\begin{equation}
P=-\frac{\Lambda}{8\pi}.\label{18}
\end{equation}%
And the mass of black holes should be interpreted as enthalpy rather than the internal energy. In this context, the Gibbs free energy can be derived through
\begin{equation}
G=H-TS=M-TS.\label{19}
\end{equation}%
After tedious calculation, we can obtain
\begin{eqnarray}
G&=&\frac{\Sigma_kr_+^{n-6}}{48\pi\alpha(r_+^2+k\alpha)^2}\Big\{(n-1)r_+^6(r_+^2+k\alpha)^2\Big[-1+\frac{(r_+^2+k\alpha)^3}{r_+^6}+\frac{12\alpha\beta^2\left(1-\frac{\Lambda}{2\beta^2}-\sqrt{1+\eta}+\frac{(n-1)\digamma(\eta)\eta}{n-2}\right)}{n(n-1)}\Big]
\nonumber
\\
&\;&-\alpha\left(\frac{r_+^4}{n-1}+\frac{2kr_+^2\alpha}{n-3}+\frac{k^2\alpha^2}{n-5}\right)\Big[(n-1)k\Big(3(n-2)r_+^4+3(n-4)k\alpha r_+^2+(n-6)k^2\alpha^2\Big)-6\Lambda r_+^6
\nonumber
\\
&\;&+12r_+^6\beta^2(1-\sqrt{1+\eta_+}\,)\Big]\Big\}
.\label{20}
\end{eqnarray}%
Imitating the approach of Refs.~\cite{Gunasekaran,Cai98}, the first law of thermodynamics in the extended phase space can be rewritten as
\begin{equation}
dM=TdS+\Phi dQ+VdP+\mathcal {A}d\alpha+\mathcal {B}d\beta,\label{21}
\end{equation}%
where $\mathcal {A}$ and $\mathcal {B}$ denote the quantities conjugated to the Lovelock coefficient and Born-Infeld parameter respectively. And they can be obtained as
\begin{eqnarray}
\mathcal {A}&=&\Big(\frac{\partial M}{\partial \alpha}\Big)_{S,Q,P,\beta}=\frac{k^2(n-1)r_+^{n-6}(3r_+^2+2k\alpha)\Sigma_k}{48\pi}
-\frac{1}{2}k(n-1)r_+^{n-5}T\Big(\frac{r_+^2}{n-3}+\frac{k\alpha}{n-5}\Big)\Sigma_k,\label{22}
\\
\mathcal {B}&=&\Big(\frac{\partial M}{\partial \beta}\Big)_{S,Q,P,\alpha}=\frac{\Sigma_kr_+^{-n}}{8n\pi \beta}\Big\{2r_+^{2n}\beta^2\Big(2-\sqrt{4+\frac{2(n-1)(n-2)q^2r_+^{2-2n}}{\beta^2}}\;\Big)
\nonumber
\\
&\;&+(n-2)(n-1)q^2r_+^2\;_2F_1\Big(\Big[\frac{1}{2},\frac{n-2}{2n-2}\Big],\Big[\frac{3n-4}{2n-2}\Big],-\frac{(n-1)(n-2)q^2}{2\beta^2r^{2n-2}}\Big)\Big\}
.\label{23}
\end{eqnarray}%
Comparing Eq. (\ref{22}) with Gauss-Bonnet black holes in Ref.~\cite{Cai98}, one may find extra terms due to the third order Lovelock gravity. Note that Eq. (\ref{21}) is limited to the case of charged topological black holes in Lovelock Born-Infeld gravity in which the second and the third order Lovelock coefficients are related via the Lovelock coefficient $\alpha$. For a general case and a nice physical interpretation of the quantity conjugated to the Lovelock coefficient, see Ref.~\cite{Kastor2}, where the Smarr relation and the first law of thermodynamics in Lovelock gravity was thoroughly investigated and it was shown that the conjugate quantity $\Psi^{(k)}$ to the Lovelock coefficient $b_k$ consists of three terms related to mass, entropy and the anti-symmetric Killing-Lovelock potential respectively.

\section{$P-V$ criticality of a limit case}
\label{Sec3}
To concentrate on the effects of the third order Lovelock gravity, we would like to investigate an interesting limit case in this section and leave the issue of nonlinear electrodynamics to be further investigated in Sec.~\ref{Sec4}.

When $\beta\rightarrow\infty$, the Born-Infeld Lagrangian reduces to the Maxwell form and $\digamma(\eta)\rightarrow1$. So one can have
\begin{equation}
g(r)\rightarrow1+\frac{3\alpha m}{r^n}+\frac{6\alpha \Lambda}{n(n-1)}-\frac{3\alpha q^2}{r^{2n-2}}.\label{24}
\end{equation}%
And the temperature for this limit case can be simplified as
\begin{equation}
T=\frac{(n-1)k[3(n-2)r_+^4+3(n-4)k\alpha r_+^2+(n-6)k^2\alpha^2]-6\Lambda r_+^6-3(n-2)(n-1)q^2r_+^{8-2n}}{12\pi(n-1)r_+(r_+^2+k\alpha)^2}.\label{25}
\end{equation}%
Substituting Eq.~(\ref{18}) into Eq.~(\ref{25}), one can find the expression for $P$ as
\begin{equation}
P=\frac{n-1}{48\pi}\Big[\frac{12\pi T}{r_+}+\frac{24k\pi \alpha T}{r_+^3}+\frac{12k^2\pi\alpha^2T}{r_+^5}+\frac{3k(2-n)}{r_+^2}+\frac{3k^2\alpha(4-n)}{r_+^4}-\frac{k^3(n-6)\alpha^2}{r_+^6}+3(n-2)q^2r_+^{2-2n}\Big].\label{26}
\end{equation}%
We can identify the specific volume $v$ as
 \begin{equation}
v=\frac{4r_+}{n-1}.\label{27}
\end{equation}%
Then Eq.~(\ref{26}) can be transformed into
\begin{equation}
P=\frac{T}{v}+\frac{32kT\alpha}{(n-1)^2v^3}+\frac{256k^2T\alpha^2}{(n-1)^4v^5}-\frac{k(n-2)}{(n-1)\pi v^2}-\frac{16k^2(n-4)\alpha}{(n-1)^3\pi v^4}-\frac{256k^3(n-6)\alpha^2}{3(n-1)^5\pi v^6}+\frac{16^{n-2}(n-2)q^2}{\pi(n-1)^{2n-3}v^{2n-2}}.\label{28}
\end{equation}%
The possible critical point should satisfy the following conditions
\begin{eqnarray}
\frac{\partial P}{\partial v}&=&0,\label{29}\\
\frac{\partial^2 P}{\partial v^2}&=&0.\label{30}
\end{eqnarray}%

Firstly, we would focus on the spherical case corresponding to $k=1$. The equation of state reads
\begin{equation}
P=\frac{T}{v}+\frac{32T\alpha}{(n-1)^2v^3}+\frac{256T\alpha^2}{(n-1)^4v^5}-\frac{(n-2)}{(n-1)\pi v^2}-\frac{16(n-4)\alpha}{(n-1)^3\pi v^4}-\frac{256(n-6)\alpha^2}{3(n-1)^5\pi v^6}+\frac{16^{n-2}(n-2)q^2}{\pi(n-1)^{2n-3}v^{2n-2}}.\label{31}
\end{equation}%

When $q=0,n=6$, Eqs.~(\ref{29}) and (\ref{30}) can be analytically solved and the corresponding physical quantities can be obtained as
\begin{equation}
T_c=\frac{1}{\pi\sqrt{5\alpha}},\;v_c=\frac{4\sqrt{\alpha}}{\sqrt{5}},\;P_c=\frac{17}{200\pi \alpha},\;\frac{P_cv_c}{T_c}=\frac{17}{50}.\label{32}
\end{equation}%
We can see clearly that the critical temperature is inversely proportional to $\sqrt{\alpha}$ while the critical specific volume is proportional to it. The critical pressure is inversely proportional to $\alpha$. However, the ratio $\frac{P_cv_c}{T_c}$ is independent of the parameter $\alpha$. Our results demonstrate again that $P-V$ criticality may exist even in the uncharged case. That may be attributed to the effect of higher derivative terms of curvature.

When $q\neq0,n=6$, one can obtain the corresponding physical quantities at the critical point as listed in Table \ref{tb1} by solving Eqs.~(\ref{29}) and (\ref{30}) numerically. From Table \ref{tb1}, one can find that there exists only one critical point for all the cases studied. And the physical quantities at the critical point $T_c, v_c,P_c$ depend on both the charge and the parameter $\alpha$ which is related to the second and the third order Lovelock coefficients. With the increasing of $\alpha$ or $q$, both $T_c$ and $P_c$ decrease while $v_c$ increases. However the ratio $\frac{P_cv_c}{T_c}$ decreases with $\alpha$ but increases with $q$.

\begin{table}[!h]
\tabcolsep 0pt
\caption{Critical values for $k=1,n=6,\beta\rightarrow\infty$}
\vspace*{-12pt}
\begin{center}
\def\temptablewidth{0.5\textwidth}
{\rule{\temptablewidth}{1pt}}
\begin{tabular*}{\temptablewidth}{@{\extracolsep{\fill}}cccccc}
$q$ & $\alpha$ & $T_c$ &$v_c$ &$P_c$ &$\frac{P_cv_c}{T_c}$ \\   \hline
     0.5  & 1 &0.14213 &       1.80992& 0.02691& 0.343  \\
       2     &1  & 0.13989&      1.97347 & 0.02553& 0.360  \\
       1     & 1  &0.14154        & 1.85884& 0.02653 & 0.348   \\
           1     & 0.5  &0.19287&        1.53461& 0.04727 & 0.376   \\
               1     & 2  &0.10062&        2.53773& 0.01351& 0.341
       \end{tabular*}
       {\rule{\temptablewidth}{1pt}}
       \end{center}
       \label{tb1}
       \end{table}

\begin{figure*}
\centerline{\subfigure[]{\label{1a}
\includegraphics[width=8cm,height=6cm]{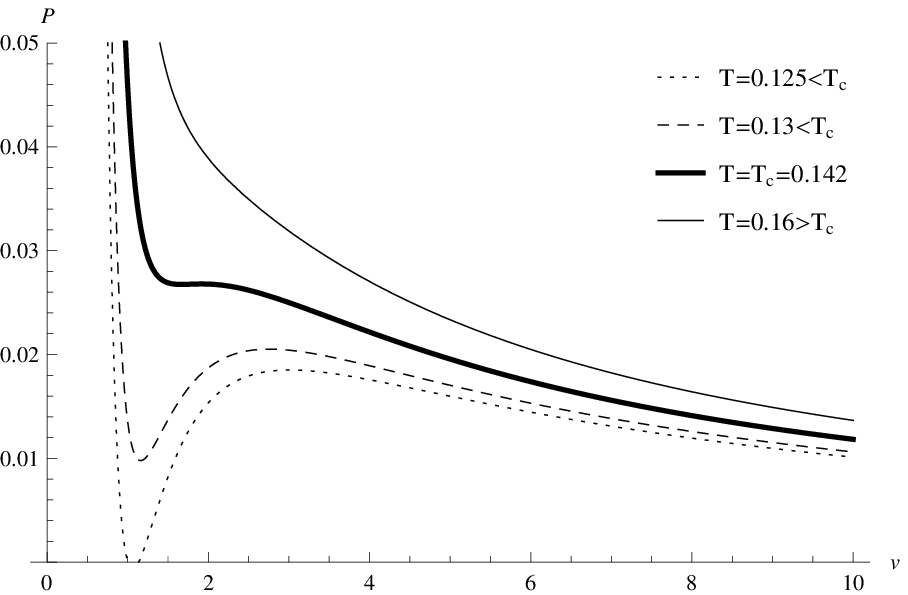}}
\subfigure[]{\label{1b}
\includegraphics[width=8cm,height=6cm]{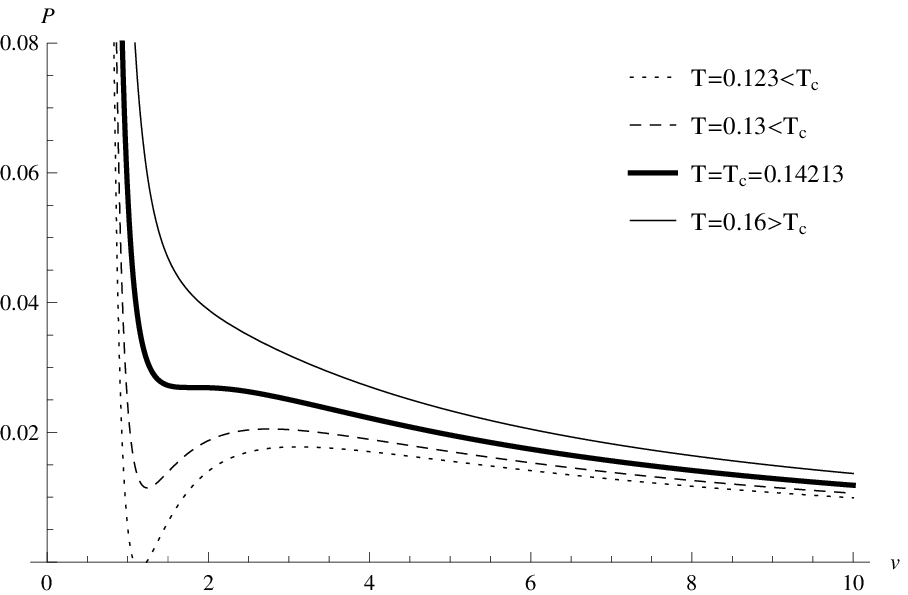}}}
\centerline{\subfigure[]{\label{1c}
\includegraphics[width=8cm,height=6cm]{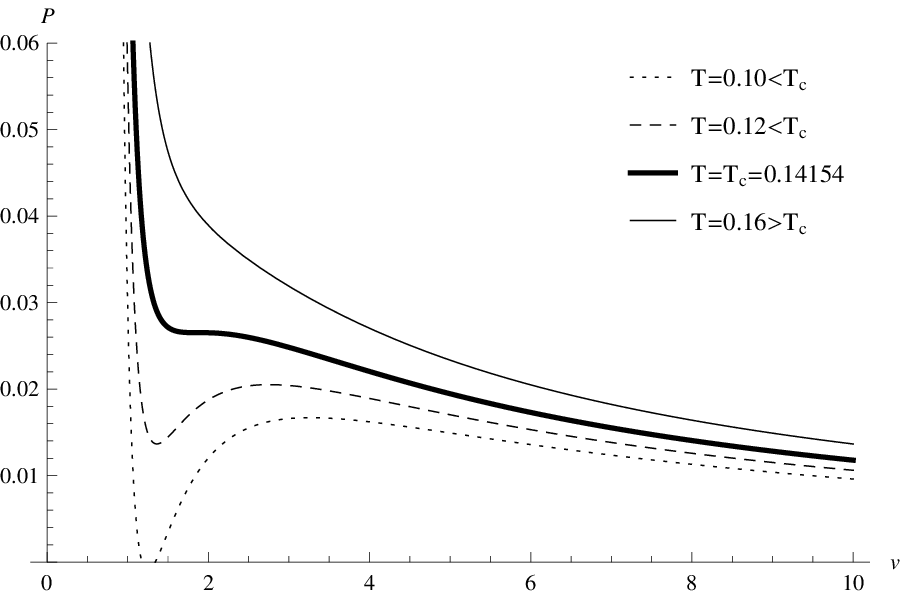}}
\subfigure[]{\label{1d}
\includegraphics[width=8cm,height=6cm]{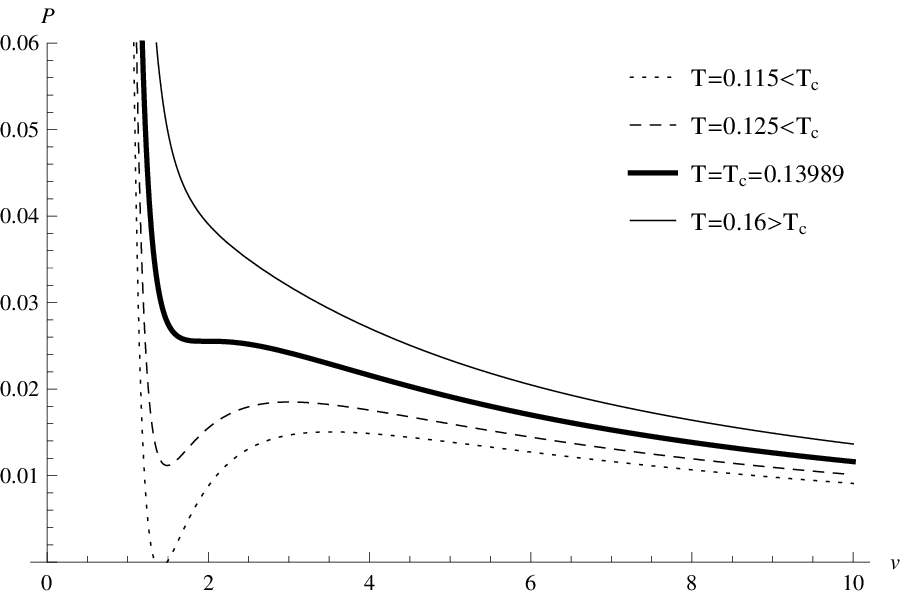}}}
\centerline{\subfigure[]{\label{1e}
\includegraphics[width=8cm,height=6cm]{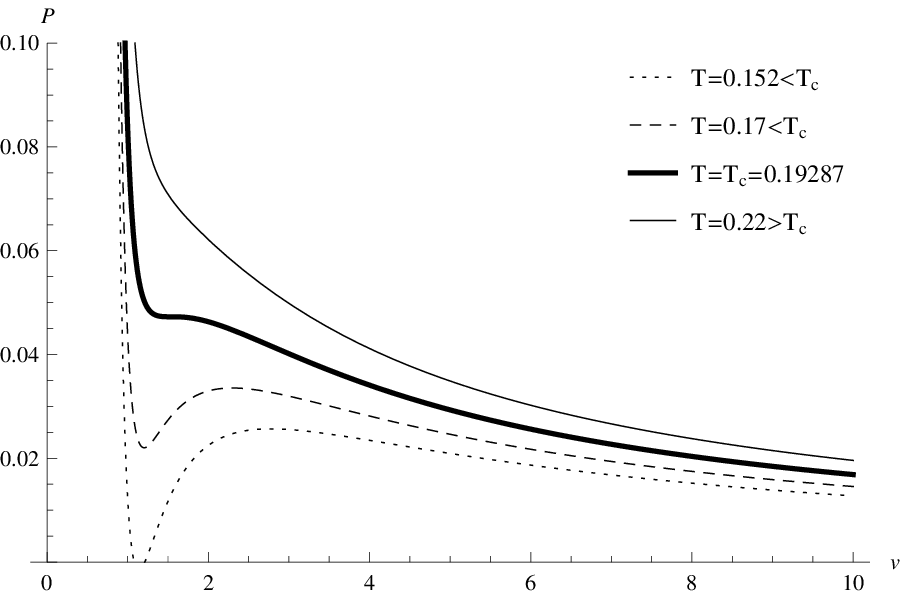}}
\subfigure[]{\label{1f}
\includegraphics[width=8cm,height=6cm]{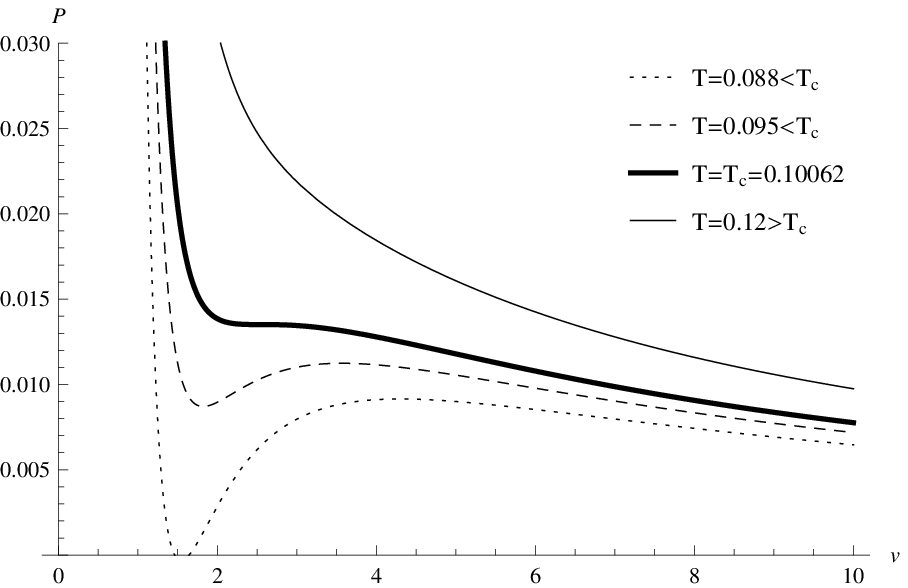}}}
 \caption{$P$ vs. $v$ for
(a)$n=6,\alpha=1,q=0$, (b) $n=6,\alpha=1,q=0.5$, (c) $n=6,\alpha=1,q=1$, (d) $n=6,\alpha=1,q=2$, (e) $n=6,\alpha=0.5,q=1$ and (f) $n=6,\alpha=2,q=1$} \label{fg1}
\end{figure*}

\begin{figure*}
\centerline{\subfigure[]{\label{2a}
\includegraphics[width=8cm,height=6cm]{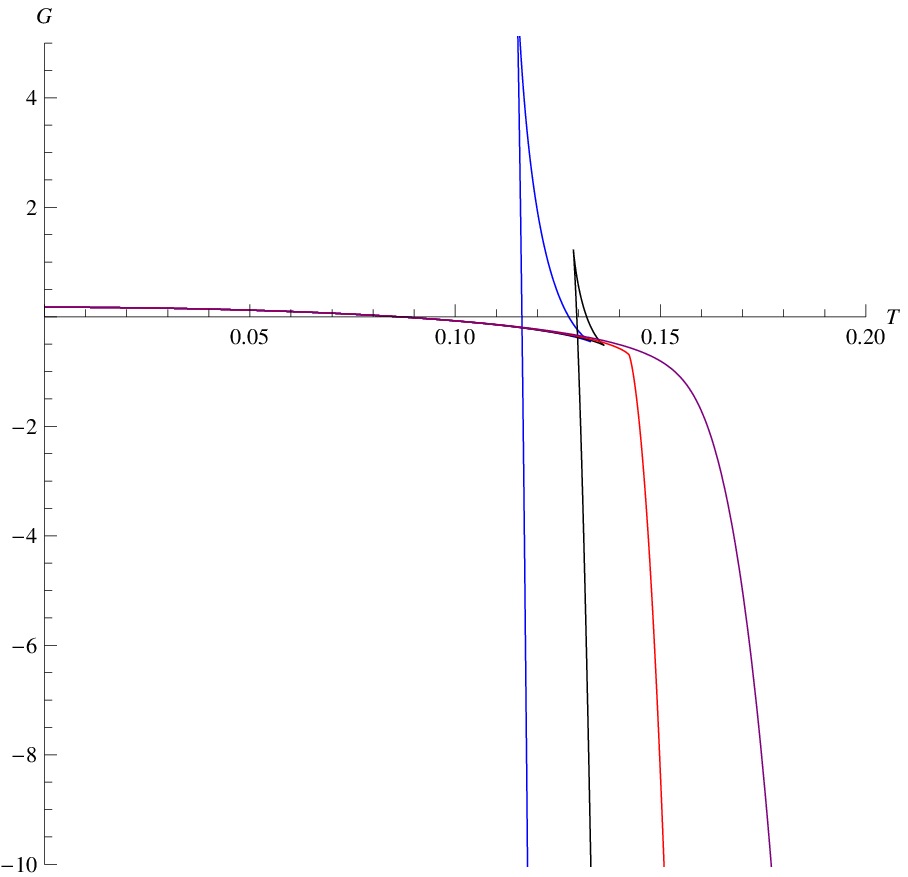}}
\subfigure[]{\label{2b}
\includegraphics[width=8cm,height=6cm]{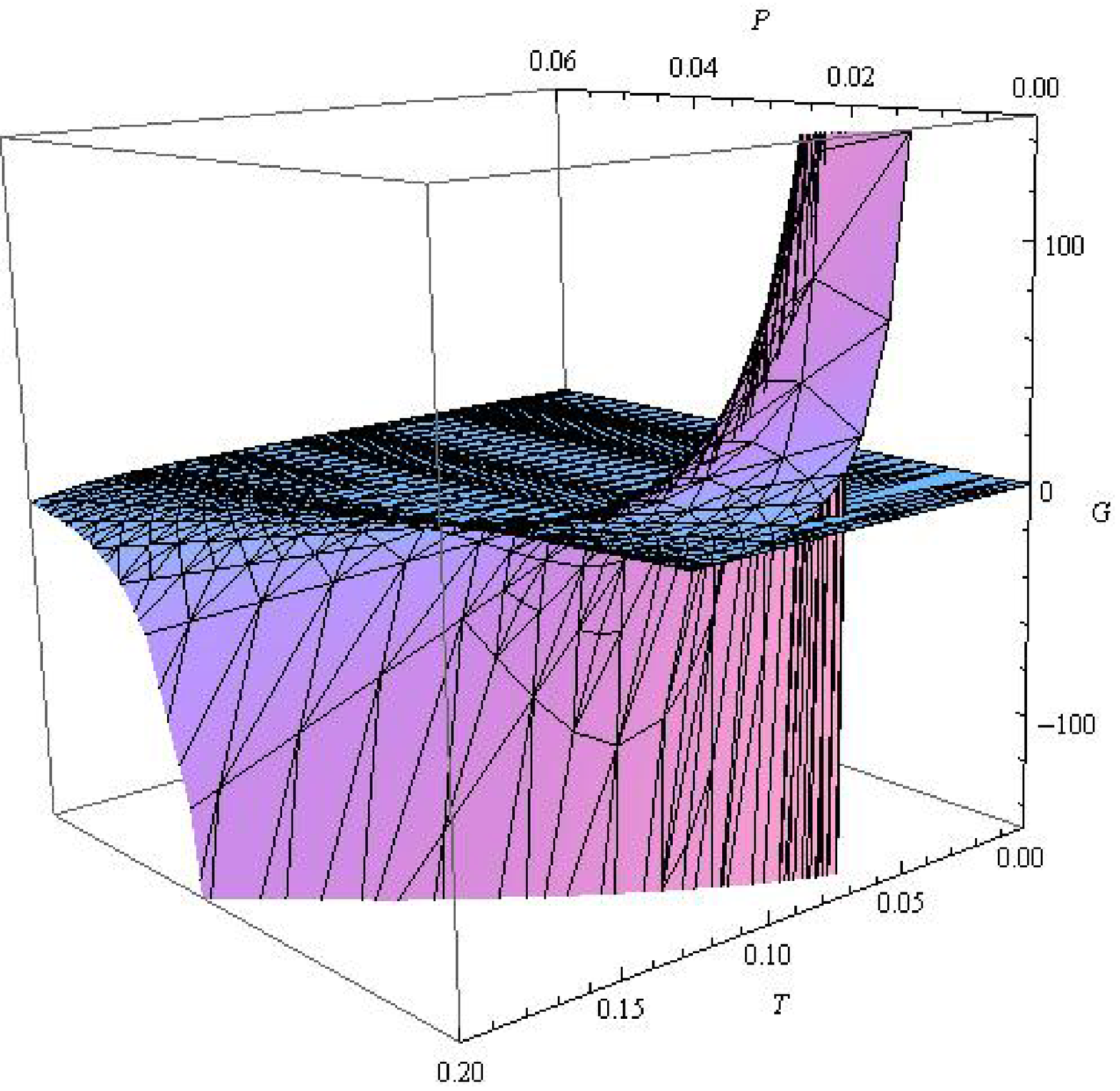}}}
 \caption{(a) $G$ vs. $T$ for $k=1, n=6,\alpha=1,q=0$, "$P=0.015<P_c$, Blue curve", "$P=0.02<P_c$, Black curve", "$P=P_c=0.02706$, Red curve", "$P=0.04>P_c$, Purple curve"  (b) $G$ vs. $P$ and $T$ for $k=1, n=6,\alpha=1,q=0$} \label{fg2}
\end{figure*}

\begin{figure*}
\centerline{\subfigure[]{\label{3a}
\includegraphics[width=8cm,height=6cm]{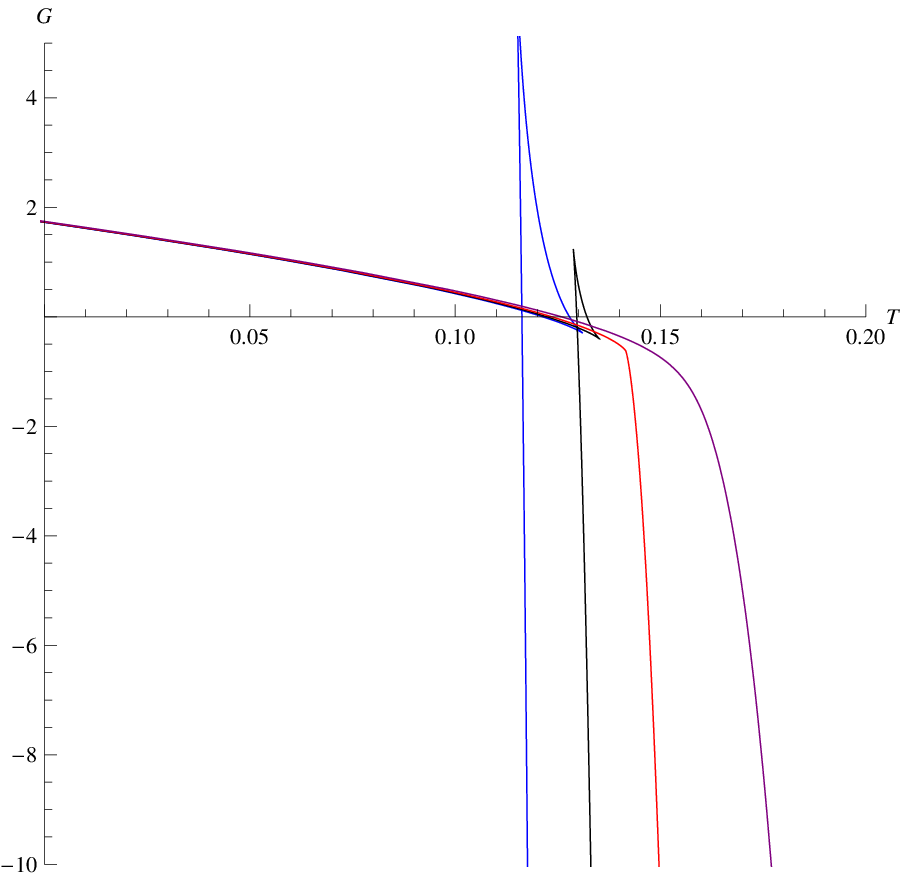}}
\subfigure[]{\label{3b}
\includegraphics[width=8cm,height=6cm]{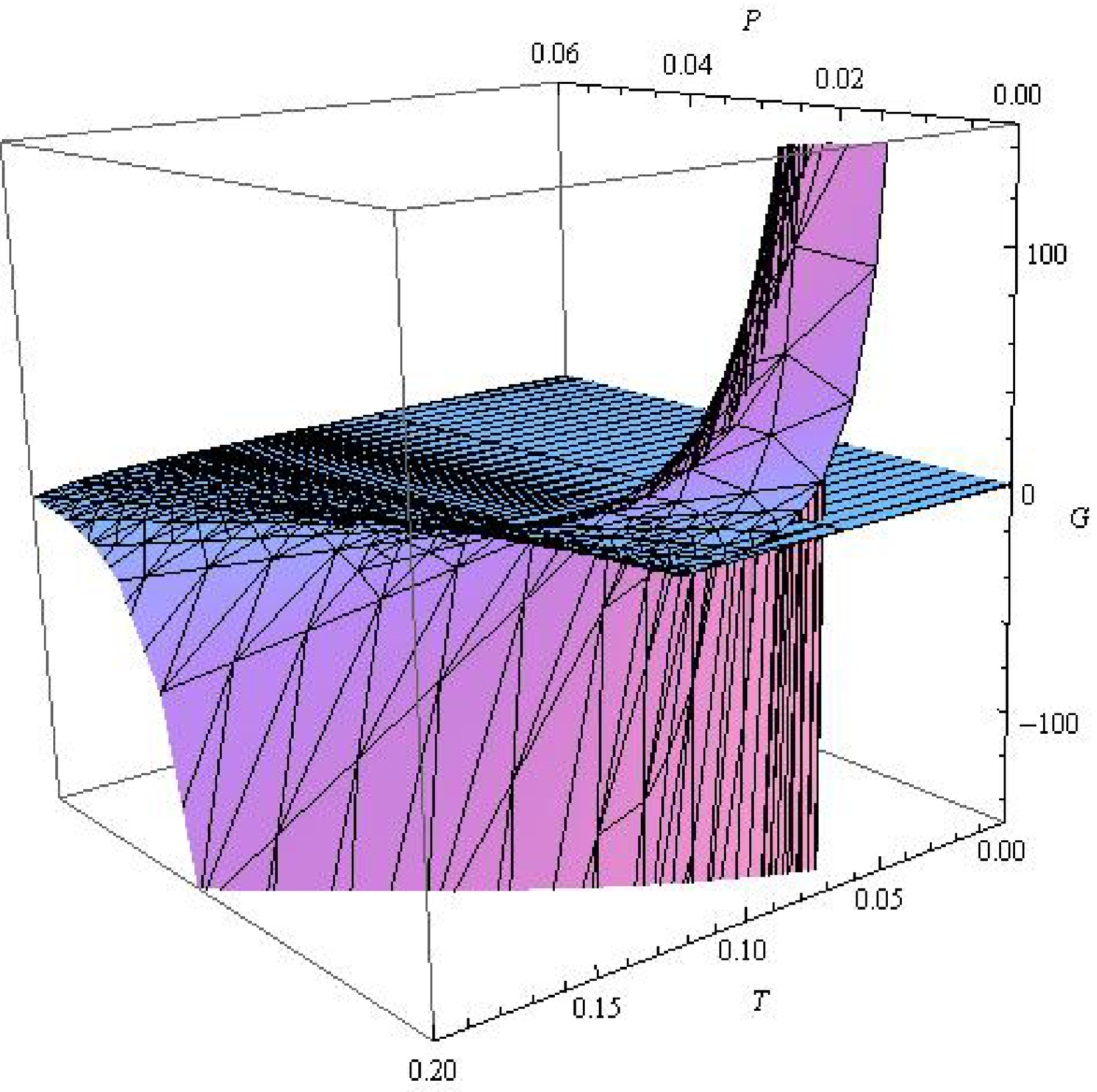}}}
 \caption{(a) $G$ vs. $T$ for $k=1, n=6,\alpha=1,q=1$, "$P=0.015<P_c$, Blue curve", "$P=0.02<P_c$, Black curve", "$P=P_c=0.02653$, Red curve", "$P=0.04>P_c$, Purple curve"  (b) $G$ vs. $P$ and $T$ for $k=1, n=6,\alpha=1,q=1$} \label{fg3}
\end{figure*}

To witness the $P-V$ criticality behavior more intuitively, we plot the $P-v$ diagram in Fig. \ref{fg1}. When the temperature is less than the critical temperature $T_c$, the isotherm can be divided into three branches. Both the large radius branch and the small radius branch are stable corresponding to a positive compression coefficient while the medium radius branch is unstable corresponding to a negative compression coefficient. The phase transition between the small black hole and the large black hole is analogous to the van der Waals liquid-gas phase transition. Figs. \ref{1a}, \ref{1b}, \ref{1c} and \ref{1d} show the impact of the charge on the $P-V$ criticality while Figs. \ref{1c}, \ref{1e}, and \ref{1f} show the effect of $\alpha$. The comparisons are in accord with the analytical results in Table \ref{tb1}. We also plot both the two-dimensional and three dimensional Gibbs free energy graph for $q=0,n=6$ in Fig. \ref{fg2} and for the case $q=1,n=6$ in Fig. \ref{fg3}. Below the critical temperature, the Gibbs free energy graphs display the classical swallow tail behavior implying the occurrence of the first order phase transition. Above the critical temperature, there is no swallow tail behavior.
%%%%%%%%%%%%%%%%%%%%%%%%%%%%%%%%%%%%%%%%%%%%%%%%%%%%%%%%%%%%%%%%%%%%%%%%%%%%%
%%%%%%%%%%%%%%%%%%%%%%%%%%%%%%%%%%%%%%%%%%%%%%%%%%%%%%%%%%%%%%%%%%%%%%%%%%%%%%%%

%%%%%%%%%%%%%%%%%%%%%%%%%%%%%%%%%%%%%%%%%%%%%%%%%%%%%%%%%%%%%%%%%%%%%%%%%%%%%
%%%%%%%%%%%%%%%%%%%%%%%%%%%%%%%%%%%%%%%%%%%%%%%%%%%%%%%%%%%%%%%%%%%%%%%%%%%%%%%%

%%%%%%%%%%%%%%%%%%%%%%%%%%%%%%%%%%%%%%%%%%%%%%%%%%%%%%%%%%%%%%%%%%%%%%%%%%%%%
%%%%%%%%%%%%%%%%%%%%%%%%%%%%%%%%%%%%%%%%%%%%%%%%%%%%%%%%%%%%%%%%%%%%%%%%%%%%%%%%

Secondly, we would discuss the $k=0$ case corresponding to Ricci flat topology. The equation of state reads
\begin{equation}
P=\frac{T}{v}+\frac{16^{n-2}(n-2)q^2}{\pi(n-1)^{2n-3}v^{2n-2}}.\label{33}
\end{equation}%
For $n=6$, utilizing Eq. (\ref{33}), one can obtain
\begin{equation}
\frac{\partial P}{\partial v}=-\frac{T}{v^2}-\frac{524288q^2}{390625\pi v^{11}},\label{34}
\end{equation}%
which is always negative for nontrivial temperature. So there would be no $P-V$ criticality for $k=0$.

Thirdly, we would investigate the $k=-1$ case corresponding to hyperbolic topology. The equation of state reads
\begin{equation}
P=\frac{T}{v}-\frac{32T\alpha}{(n-1)^2v^3}+\frac{256T\alpha^2}{(n-1)^4v^5}+\frac{(n-2)}{(n-1)\pi v^2}-\frac{16(n-4)\alpha}{(n-1)^3\pi v^4}+\frac{256(n-6)\alpha^2}{3(n-1)^5\pi v^6}+\frac{16^{n-2}(n-2)q^2}{\pi(n-1)^{2n-3}v^{2n-2}}.\label{35}
\end{equation}%
Similarly, when $q=0,n=6$, Eqs.~(\ref{29}) and (\ref{30}) can be analytically solved and the corresponding physical quantities can be obtained as
\begin{equation}
T_c=\frac{1}{2\pi\sqrt{\alpha}},\;v_c=\frac{4\sqrt{\alpha}}{5},\;P_c=\frac{5}{8\pi \alpha},\;\frac{P_cv_c}{T_c}=1.\label{36}
\end{equation}%
When $q\neq0,n=6$, one can obtain the numerical solutions of Eqs.~(\ref{29}) and (\ref{30}) as listed in Table \ref{tb2}. These results are quite different from those in former literature which demonstrated that $P-V$ criticality only exists in the $k=1$ case for topological black holes in both Einstein gravity and Gauss-Bonnet gravity~\cite{Kubiznak,Cai98}.
\begin{table}[!h]
\tabcolsep 0pt
\caption{Critical values for $k=-1,n=6,\beta\rightarrow\infty$}
\vspace*{-12pt}
\begin{center}
\def\temptablewidth{0.5\textwidth}
{\rule{\temptablewidth}{1pt}}
\begin{tabular*}{\temptablewidth}{@{\extracolsep{\fill}}cccccc}
$q$ & $\alpha$ & $T_c$ &$v_c$ &$P_c$ &$\frac{P_cv_c}{T_c}$ \\   \hline
     0.5  & 1 &0.33836 &       1.07752& 0.22718& 0.723  \\
       2     &1  & 0.72658&      1.31811 & 0.35029& 0.635  \\
       1     & 1  &0.46900        & 1.19335& 0.26507 & 0.674   \\
           1     & 0.5  &1.88727&        1.01514& 1.12928 & 0.607   \\
               1     & 2  &0.18669&        1.38663& 0.10503& 0.780
       \end{tabular*}
       {\rule{\temptablewidth}{1pt}}
       \end{center}
       \label{tb2}
       \end{table}

\begin{figure*}
\centerline{\subfigure[]{\label{4a}
\includegraphics[width=8cm,height=6cm]{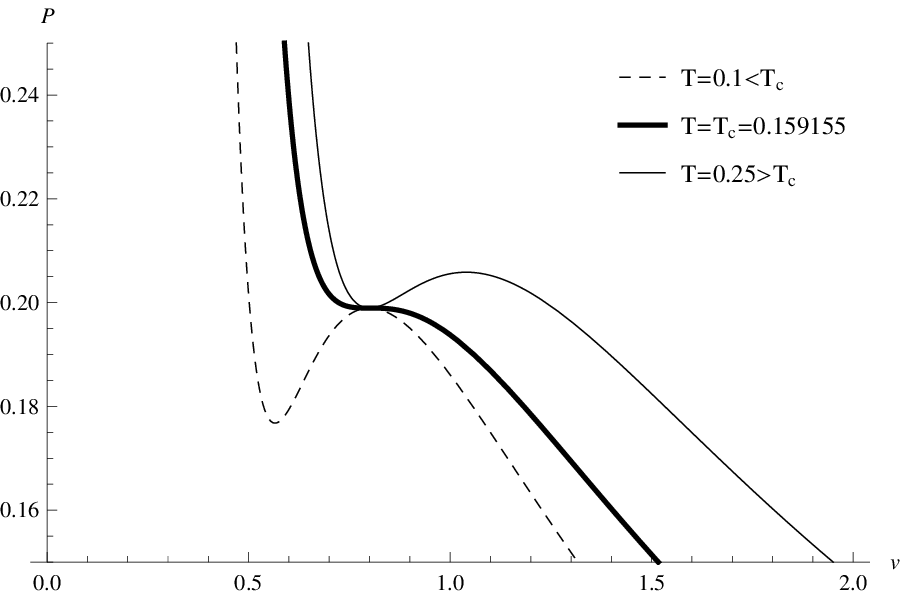}}
\subfigure[]{\label{4b}
\includegraphics[width=8cm,height=6cm]{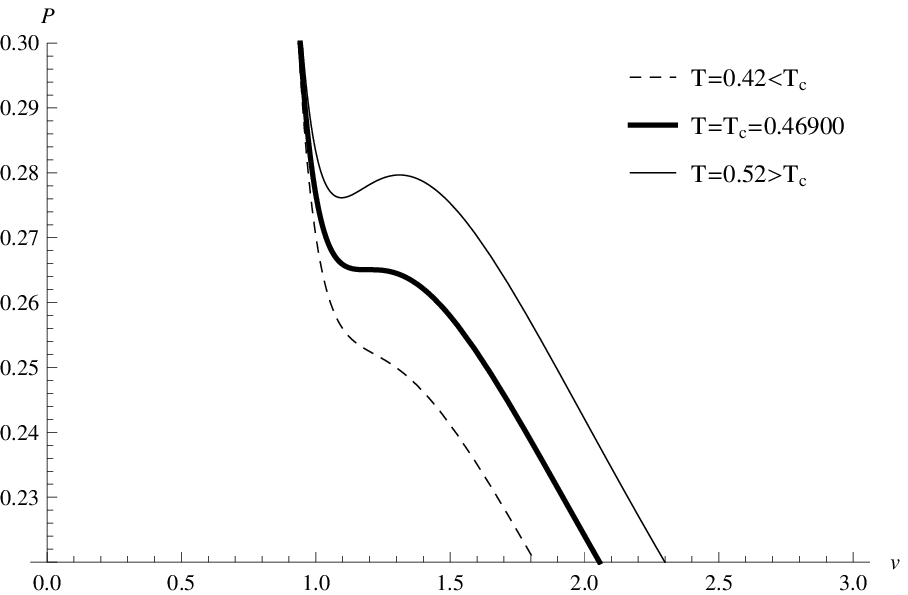}}}
 \caption{$P$ vs. $v$ for
(a) $k=-1, n=6,\alpha=1,q=0$, (b) $k=-1, n=6,\alpha=1,q=1$} \label{fg4}
\end{figure*}

\begin{figure*}
\centerline{\subfigure[]{\label{5a}
\includegraphics[width=8cm,height=6cm]{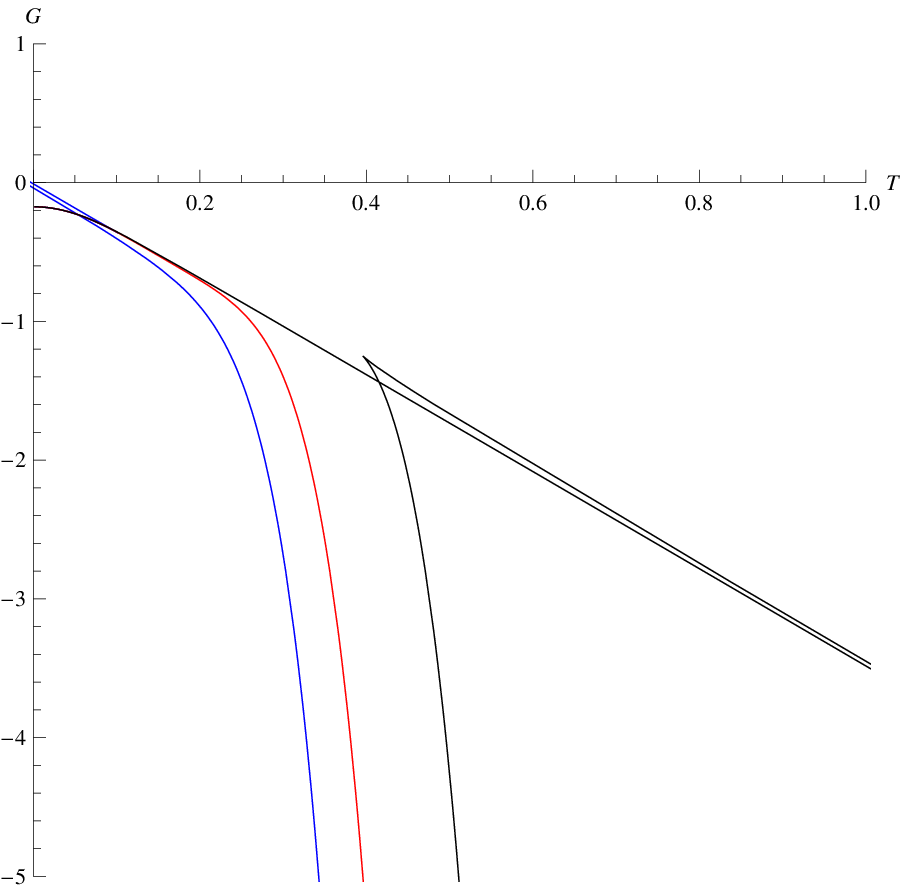}}
\subfigure[]{\label{5b}
\includegraphics[width=8cm,height=6cm]{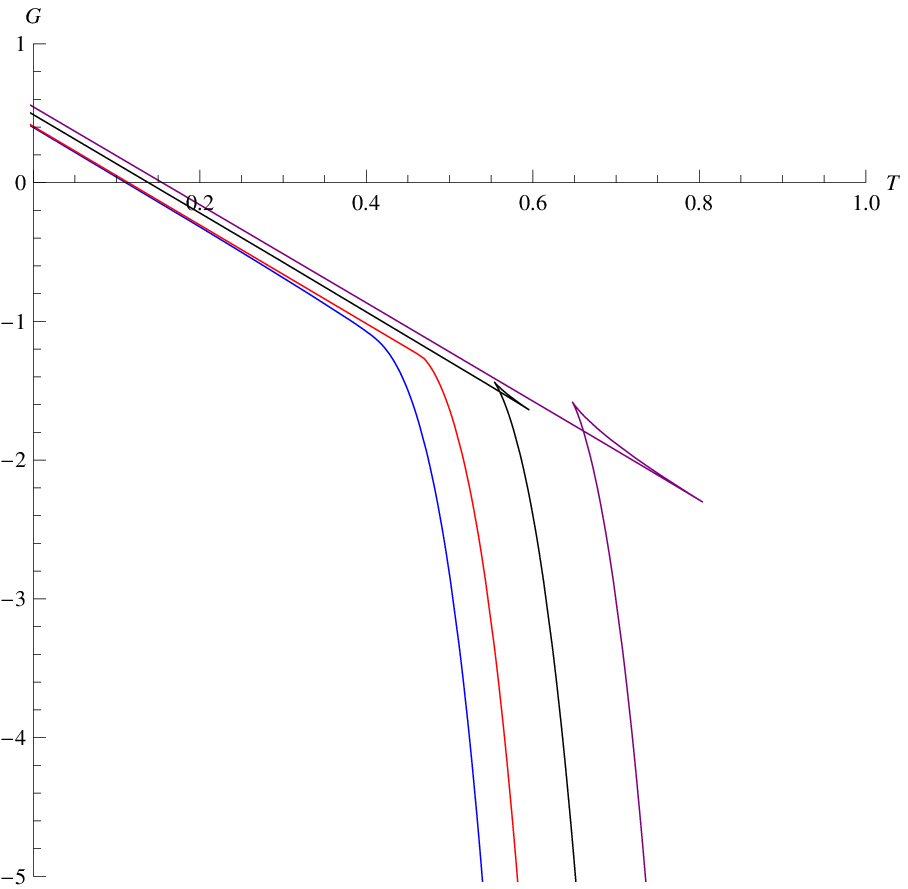}}}
 \caption{$G$ vs. $T$ for
(a)$k=-1, n=6,\alpha=1,q=0$, "$P=0.15<P_c$, Blue curve", "$P=P_c=0.19894$, Red curve", "$P=0.24>P_c$, Black curve" (b) $k=-1, n=6,\alpha=1,q=1$,"$P=0.25<P_c$, Blue curve", "$P=P_c=0.26507$, Red curve", "$P=0.29>P_c$, Black curve","$P=0.32>P_c$, Purple curve"} \label{fg5}
\end{figure*}

To gain an intuitive picture, we plot the $P-v$ diagram in Fig. \ref{fg4}, which shows strange behaviors different from van der Waals liquid-gas phase transition. The isotherm at the critical temperature is quite similar to the van der Waals liquid-gas system. However, for the uncharged case in Fig. \ref{4a}, the isotherm below or above the critical temperature both behave as the coexistence phase which is similar to the behaviors of van der Waals liquid-gas system below the critical temperature.  For the charged case in Fig. \ref{4b}, the "phase transition" picture is quite the reverse of van der Waals liquid-gas phase transition. Above the critical temperature the behavior is "van der Waals like" while the behavior is "ideal gas like" below the critical temperature. This process is achieved by lowering the temperature rather than increasing the temperature. We also plot the Gibbs free energy in Fig. \ref{fg5} and "swallow tail" behavior can be observed.
%%%%%%%%%%%%%%%%%%%%%%%%%%%%%%%%%%%%%%%%%%%%%%%%%%%%%%%%%%%%%%%%%%%%%%%%%%%%%
%%%%%%%%%%%%%%%%%%%%%%%%%%%%%%%%%%%%%%%%%%%%%%%%%%%%%%%%%%%%%%%%%%%%%%%%%%%%%%%%

The results above are so strange that motivates us to check whether they are physical. The non-negative definiteness of entropy demands that
\begin{equation}
\frac{r_+^4}{n-1}+\frac{2kr_+^2\alpha}{n-3}+\frac{k^2\alpha^2}{n-5}\geq0.\label{37}
\end{equation}%
In fact, when $n=6$, the L.H.S. of the above inequality can be obtained by utilizing Eq. (\ref{27}) as
\begin{equation}
\frac{125v^4}{256}-\frac{25v^2\alpha}{24}+\alpha^2.\label{38}
\end{equation}%
Denoting $v^2$ as $x$, one can consider the equation
\begin{equation}
\frac{125x^2}{256}-\frac{25\alpha x}{24}+\alpha^2=0,\label{39}
\end{equation}%
with the discriminant as
\begin{equation}
\Delta=\left(\frac{25\alpha}{24}\right)^2-4\times\alpha^2\times\frac{125}{256}=-\frac{125\alpha^2}{144}.\label{40}
\end{equation}%
Note that for any nontrivial value of $\alpha$, the discriminant of Eq. (\ref{39}) is always negative, implying that the values of entropy are always positive for any specific volume $v$.
%%%%%%%%%%%%%%%%%%%%%%%%%%%%%%%%%%%%%%%%%%%%%%%%%%%%%%%%%%%%%%%%%%%%%%%%%%%%%
%%%%%%%%%%%%%%%%%%%%%%%%%%%%%%%%%%%%%%%%%%%%%%%%%%%%%%%%%%%%%%%%%%%%%%%%%%%%%%%%
\section{Inclusion of the nonlinear electrodynamics}
\label{Sec4}
In this section, we would like to take into account the effect of non-linear electrodynamics to complete the analysis of topological AdS black holes in Lovelock-Born-Infeld gravity.

Utilizing Eqs. (\ref{13}) and (\ref{18}), Eq. (\ref{14}) can be rewritten as
\begin{eqnarray}
P&=&\frac{T}{v}+\frac{32kT\alpha}{(n-1)^2v^3}+\frac{256k^2T\alpha^2}{(n-1)^4v^5}-\frac{k(n-2)}{(n-1)\pi v^2}-\frac{16k^2(n-4)\alpha}{(n-1)^3\pi v^4}-\frac{256k^3(n-6)\alpha^2}{3(n-1)^5\pi v^6}
\nonumber
\\
&\;&-\frac{\beta^2}{4\pi}\left\{1-\sqrt{1+\frac{2^{4n-5}(n-2)(n-1)q^2[(n-1)v]^{2-2n}}{\beta^2}}\right\}.\label{41}
\end{eqnarray}%

\begin{figure*}
\centerline{\subfigure[]{\label{6a}
\includegraphics[width=8cm,height=6cm]{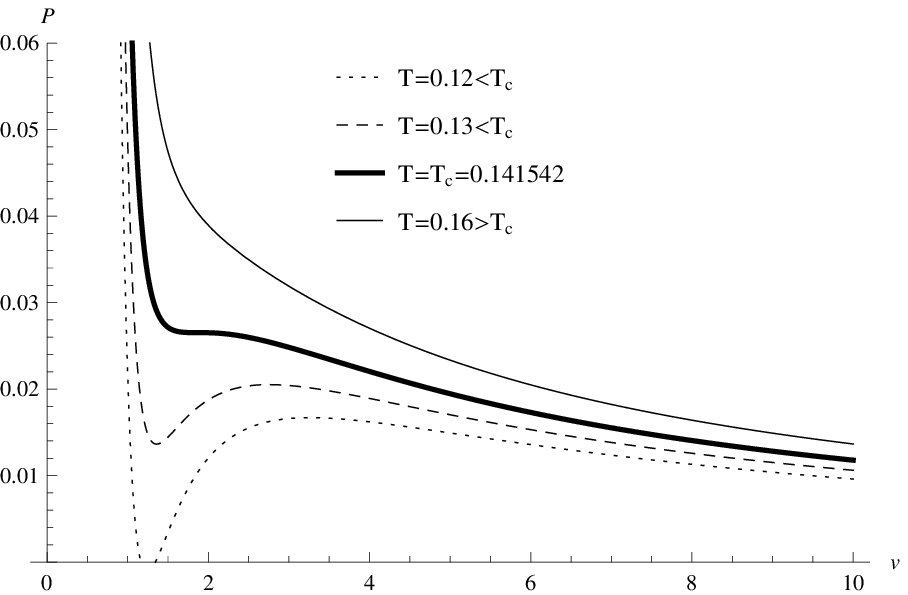}}
\subfigure[]{\label{6b}
\includegraphics[width=8cm,height=6cm]{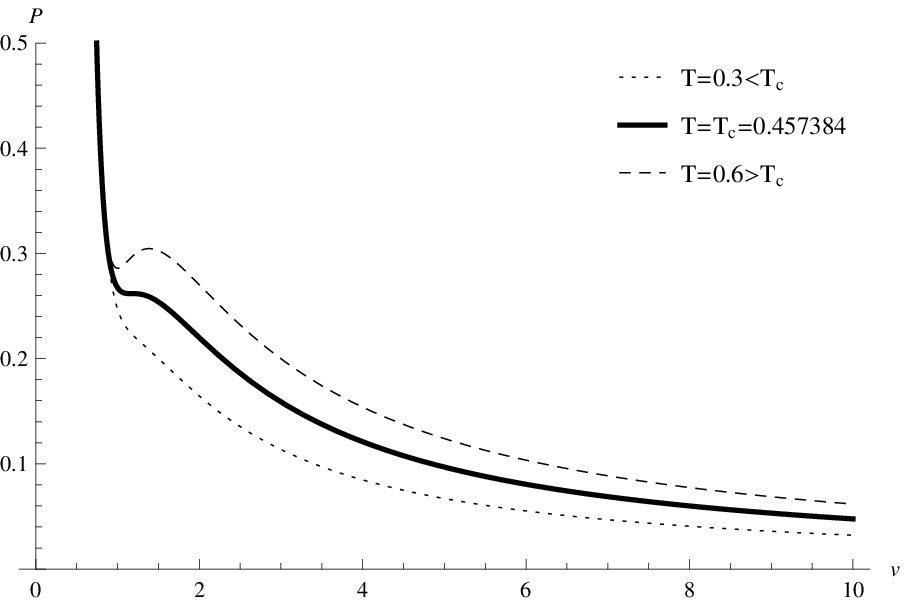}}}
 \caption{$P$ vs. $v$ for
(a)$k=1,n=6,\alpha=1,\beta=1,q=1$ and (b) $k=-1,n=6,\alpha=1,\beta=1,q=1$} \label{fg6}
\end{figure*}

\begin{figure*}
\centerline{\subfigure[]{\label{7a}
\includegraphics[width=8cm,height=6cm]{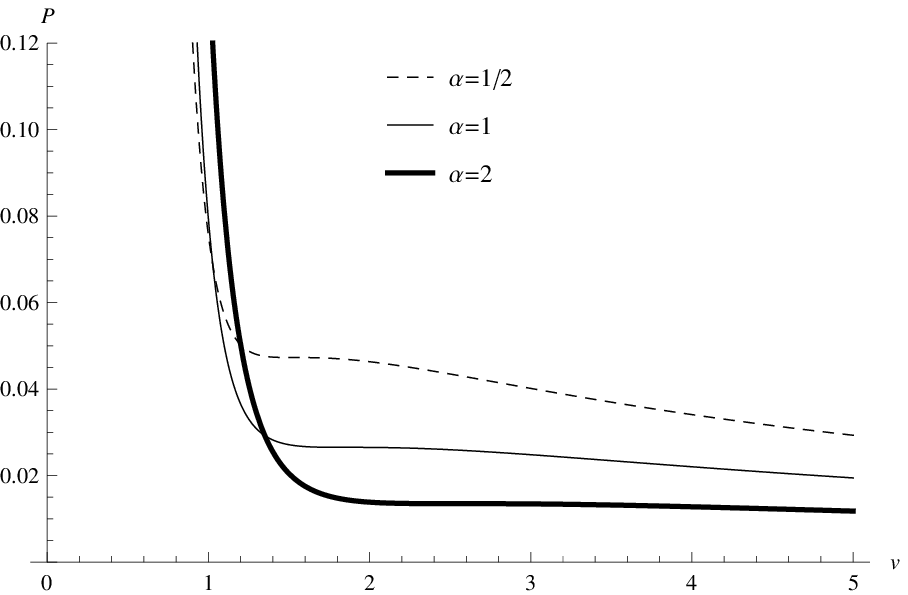}}
\subfigure[]{\label{7b}
\includegraphics[width=8cm,height=6cm]{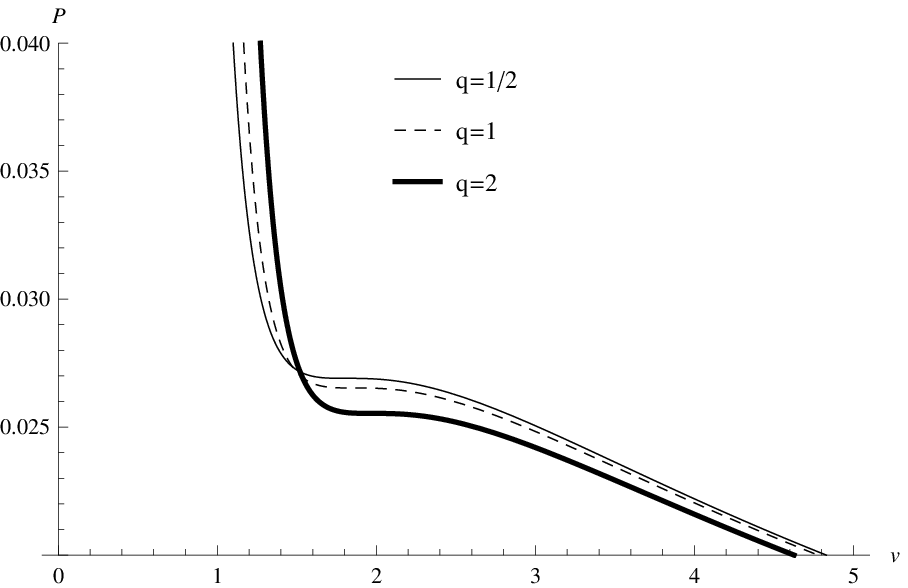}}}
 \caption{Isotherm at the critical temperature for
(a)$k=1,n=6,\beta=1,q=1$ and (b) $k=1,n=6,\beta=1,\alpha=1$} \label{fg7}
\end{figure*}

Similarly, we would discuss the $k=1$ case corresponding to spherical topology first. The equation of state reads
\begin{eqnarray}
P&=&\frac{T}{v}+\frac{32T\alpha}{(n-1)^2v^3}+\frac{256T\alpha^2}{(n-1)^4v^5}-\frac{(n-2)}{(n-1)\pi v^2}-\frac{16(n-4)\alpha}{(n-1)^3\pi v^4}-\frac{256(n-6)\alpha^2}{3(n-1)^5\pi v^6}
\nonumber
\\
&\;&-\frac{\beta^2}{4\pi}\left\{1-\sqrt{1+\frac{2^{4n-5}(n-2)(n-1)q^2[(n-1)v]^{2-2n}}{\beta^2}}\right\}.\label{42}
\end{eqnarray}%
One can obtain the corresponding physical quantities at the critical point as listed in Table \ref{tb3} by solving Eqs.~(\ref{29}) and (\ref{30}) for the case $n=6$ numerically. As is shown, the physical quantities at the critical point $T_c, v_c,P_c$ depend on the charge, the Lovelock coefficient $\alpha$ and the Born-Infeld parameter $\beta$. With the increasing of $\alpha$ or $q$, both $T_c$ and $P_c$ decrease while $v_c$ increases. However the ratio $\frac{P_cv_c}{T_c}$ decreases with $\alpha$ but increases with $q$. These observations are similar to the limit case $\beta\rightarrow\infty$. With the increasing of $\beta$, $T_c$, $P_c$ decrease while $v_c$ and the ratio $\frac{P_cv_c}{T_c}$ increase. However, only slight differences can be observed concerning the impact of nonlinear electrodynamics. That may be attributed to the parameter region we choose. Readers who are interested in the "Schwarzschild like" behavior of Born-Infeld black holes can read the interesting paper Ref.~\cite{Gunasekaran}. For an intuitive understanding, we plot the $P-v$ diagram in Fig. \ref{6a} and show the effect of the parameter $q$ and $\alpha$ in Fig. \ref{fg7}.
\begin{table}[!h]
\tabcolsep 0pt
\caption{Critical values for different dimensions for $k=1,n=6$}
\vspace*{-12pt}
\begin{center}
\def\temptablewidth{0.5\textwidth}
{\rule{\temptablewidth}{1pt}}
\begin{tabular*}{\temptablewidth}{@{\extracolsep{\fill}}cccccccc}
$\beta$ & $q$ & $\alpha$ &$T_c$ &$v_c$ &$P_c$ &$\frac{P_cv_c}{T_c}$ \\   \hline
          10  & 1 &1 &0.141541&       1.85884& 0.026528 & 0.34839 \\
    0.5  & 1 &1 &0.141545&       1.85829& 0.026531 & 0.34832  \\
     1  & 1  &1 &0.141542&        1.85871& 0.026529 & 0.34838 \\
    1  & 0.5  &1 &0.142126&       1.80991& 0.02691 & 0.343 \\
  1  & 2 &1  &0.139898&        1.97286& 0.02554 & 0.360 \\
     1  & 1  &0.5 &0.192905&        1.53258& 0.04730 & 0.376 \\
      1  & 1  &2 &0.100617&        2.53773& 0.01351 & 0.341
       \end{tabular*}
       {\rule{\temptablewidth}{1pt}}
       \end{center}
       \label{tb3}
       \end{table}
%%%%%%%%%%%%%%%%%%%%%%%%%%%%%%%%%%%%%%%%%%%%%%%%%%%%%%%%%%%%%%%%%%%%%%%%%%%%%
%%%%%%%%%%%%%%%%%%%%%%%%%%%%%%%%%%%%%%%%%%%%%%%%%%%%%%%%%%%%%%%%%%%%%%%%%%%%%%%%

%%%%%%%%%%%%%%%%%%%%%%%%%%%%%%%%%%%%%%%%%%%%%%%%%%%%%%%%%%%%%%%%%%%%%%%%%%%%%
%%%%%%%%%%%%%%%%%%%%%%%%%%%%%%%%%%%%%%%%%%%%%%%%%%%%%%%%%%%%%%%%%%%%%%%%%%%%%%%%

Secondly, we would discuss the $k=0$ case corresponding to Ricci flat topology. The equation of state reads
\begin{equation}
P=\frac{T}{v}-\frac{\beta^2}{4\pi}\left\{1-\sqrt{1+\frac{2^{4n-5}(n-2)(n-1)q^2[(n-1)v]^{2-2n}}{\beta^2}}\right\}.\label{43}
\end{equation}%
There would be no $P-V$ criticality because $P$ monotonically decreases with $v$.

Thirdly, we would discuss the $k=-1$ case corresponding to hyperbolic topology. The equation of state reads
\begin{eqnarray}
P&=&\frac{T}{v}-\frac{32T\alpha}{(n-1)^2v^3}+\frac{256T\alpha^2}{(n-1)^4v^5}+\frac{(n-2)}{(n-1)\pi v^2}-\frac{16(n-4)\alpha}{(n-1)^3\pi v^4}+\frac{256(n-6)\alpha^2}{3(n-1)^5\pi v^6}
\nonumber
\\
&\;&-\frac{\beta^2}{4\pi}\left\{1-\sqrt{1+\frac{2^{4n-5}(n-2)(n-1)q^2[(n-1)v]^{2-2n}}{\beta^2}}\right\}.\label{44}
\end{eqnarray}%
Numerical solutions of Eqs.~(\ref{29}) and (\ref{30}) are listed in Table \ref{tb4} and we also plot the $P-v$ diagram in Fig. \ref{6b}, in which similar strange behavior is also observed. Note that the entropy analysis also holds because the entropy in Eq.~(\ref{17}) is independent of $\beta$. So we would not repeat the analysis here.
\begin{table}[!h]
\tabcolsep 0pt
\caption{Critical values for different dimensions for $k=-1,n=6$}
\vspace*{-12pt}
\begin{center}
\def\temptablewidth{0.5\textwidth}
{\rule{\temptablewidth}{1pt}}
\begin{tabular*}{\temptablewidth}{@{\extracolsep{\fill}}cccccccc}
$\beta$ & $q$ & $\alpha$ &$T_c$ &$v_c$ &$P_c$ &$\frac{P_cv_c}{T_c}$ \\   \hline
          10  & 1 &1 &0.468887&       1.19315& 0.26504 & 0.674 \\
    0.5  & 1 &1 &0.424488&       1.11470& 0.25296& 0.664  \\
    1  & 1  &1 &0.457384&        1.17317& 0.26182& 0.672\\
    1  & 0.5  &1 &0.333977&       1.06663&0.22621 & 0.722 \\
  1  & 2 &1  &0.690585&       1.28270& 0.33877 & 0.629 \\
     1  & 1  &0.5 &1.498501&        0.92425& 0.94270 & 0.581 \\
      1  & 1  &2 &0.186104&       1.38311& 0.10496 & 0.780
       \end{tabular*}
       {\rule{\temptablewidth}{1pt}}
       \end{center}
       \label{tb4}
       \end{table}

\section{Conclusions}
\label{Sec5}
   Till now, the topological AdS black holes in Lovelock-Born-Infeld gravity are investigated in the extended phase space. The black hole solutions are reviewed while their thermodynamics is further explored in the extended phase space. We calculate the entropy by integration and find that the result in former literature~\cite{Decheng2} was incomplete. Treating the cosmological constant as pressure, we rewrite the first law of thermodynamics for the specific case in which the second order and the third order Lovelock coefficients are related by the Lovelock coefficient $\alpha$. The quantity conjugated to Lovelock coefficient and the Born-Infeld parameter respectively are calculated. Comparing our results of the above quantities with those in former literature of Gauss-Bonnet black holes~\cite{Cai98}, we find that there exist extra terms due to the third order Lovelock gravity. In order to make the phase transition clearer, the Gibbs free energy is also calculated.

   To figure out the effect of the third order Lovelock gravity on the $P-V$ criticality, a detailed analysis of the limit case $\beta\rightarrow\infty$ has been performed. Since the entropy is convergent only when $n>5$, our investigation is carried out in the case of $n=6$, corresponding to the seven-dimensional black holes. It is shown that for the spherical topology, $P-V$ criticality exists even when $q=0$. The critical physical quantities can be analytically solved and they vary with the parameter $\alpha$. However, the ratio of $\frac{P_cv_c}{T_c}$ is independent of the parameter $\alpha$. Our results demonstrate again that the charge is not the indispensable condition of $P-V$ criticality. It may be attributed to the effect of higher derivative terms of curvature because similar phenomenon was also found for Gauss-Bonnet black holes~\cite{Cai98}. For $q\neq0$, it is shown that the physical quantities at the critical point $T_c, v_c,P_c$ depends on both the charge and the parameter $\alpha$. With the increasing of $\alpha$ or $q$, both $T_c$ and $P_c$ decrease while $v_c$ increases. However the ratio $\frac{P_cv_c}{T_c}$ decreases with $\alpha$ but increases with $q$. Similar behaviors as van der Waals liquid-gas phase transition can be observed in the $P-v$ diagram and the classical swallow tail behaviors can be observed in both the two-dimensional and three-dimensional graph of Gibbs free energy. These observations indicate that phase transition between small black holes and large black holes take place when $k=1$. For $k=0$, no critical point can be found and there would be no $P-V$ criticality. Interesting findings occur in the case $k=-1$, in which positive solutions of critical points are found for both the uncharged and charged case. However, the $P-v$ diagram is very strange. For the uncharged case, the isotherms below or above the critical temperature both behave as the coexistence phase which is similar to the behaviors of van der Waals liquid-gas system below the critical temperature.  For the charged case, the "phase transition" picture is quite the reverse of van der Waals liquid-gas phase transition. Above the critical temperature the behavior is "van der Waals like" while the behavior is "ideal gas like" below the critical temperature. This process is achieved by lowering the temperature rather than increasing the temperature. To check whether these findings are physical, we perform analysis on the non-negative definiteness condition of entropy. It is shown that for any nontrivial value of $\alpha$, the entropy is always positive for any specific volume $v$. We relate the findings in the case $k=-1$ with the peculiar property of the third order Lovelock gravity. Because the entropy in the third order Lovelock gravity consists of extra terms which is absent in the Gauss-Bonnet black holes, which makes the critical points satisfy the constraint of non-negative definiteness condition of entropy. We also check the Gibbs free energy graph and
"swallow tail" behavior can be observed.

   Moreover, the effect of nonlinear electrodynamics is included in our work. Similar observations are made as the limit case $\beta\rightarrow\infty$ and only slight differences can be observed when we choose different values of $\beta$. That may be attributed to the parameter region we choose. More interesting findings concerning the "Schwarzschild like" behaviors can be found in the former literature~\cite{Gunasekaran} and we would not repeat them here because our main motivation is to investigate the impact of the third order Lovelock gravity on the $P-V$ criticality in the extend phase space.

 \section*{Acknowledgements}
This research is supported by the National Natural Science
Foundation of China (Grant Nos.11235003, 11175019, 11178007). It is
also supported by \textquotedblleft Thousand Hundred
Ten\textquotedblright \,Project of Guangdong Province and Natural Science Foundation of Zhanjiang Normal University under
Grant No. QL1104.

\end{document}